\documentclass[aps,prc,twocolumn,superscriptaddress,showpacs]{revtex4}

\newcommand{\PL}[3]{Phys. Lett. {\bf #1}, #2 (#3)}
\newcommand{\PRL}[3]{Phys. Rev. Lett. {\bf #1}, #2 (#3)}
\newcommand{\NP}[3]{Nucl. Phys. {\bf #1}, #2 (#3)}
\newcommand{\PR}[3]{Phys. Rev. {\bf #1}, #2 (#3)}
\newcommand{\PTP}[3]{Prog. Theor. Phys. {\bf #1}, #2 (#3)}

\newcommand{\Km}{K$^-$}

\usepackage{graphicx}

\begin{document}


\title{
Kaonic nuclei studied based on a new framework of \\
Antisymmetric Molecular Dynamics
}


\author{Akinobu Dot$\acute{\rm e}$}
\affiliation{Institute of Particle and Nuclear Studies, KEK, Tsukuba, Ibaraki
305-0801, Japan}

\author{Hisashi Horiuchi}
\affiliation{Department of Physics, Kyoto University, Kyoto 606-8502, Japan}

\author{Yoshinori Akaishi}
\affiliation{Institute of Particle and Nuclear Studies, KEK, Tsukuba, Ibaraki
305-0801, Japan}

\author{Toshimitsu Yamazaki}
\affiliation{RI Beam Science Laboratory, RIKEN, Wako,
Saitama 351-0198, Japan}


\date{\today}

\begin{abstract}
We have developed a new framework of Antisymmetrized Molecular Dynamics (AMD), 
to adequately treat the $I=0$ \={K}N interaction, which is essential 
to study kaonic nuclei. 
The improved points are 1) pK$^-$/n\={K}$^0$ mixing and 2) total spin and 
isospin projections. 
These improvements enable us to investigate various kaonic nuclei 
(ppnK$^-$, pppK$^-$, pppnK$^-$, $^6$BeK$^-$ and $^9$BK$^-$) systematically.  
We have found that they are deeply bound and extremely dense 
with a variety of shapes.
\end{abstract}


\maketitle


\section{Introduction}

Recently, it was proposed that 
a K$^-$ meson can be deeply bound in light nuclei as a discrete state, 
such as $^3$He+K$^-$, $^4$He+K$^-$ and $^8$Be+K$^-$, 
where the K$^-$ meson makes the nucleus shrink drastically to form 
a dense state \cite{Akaishi-Yamazaki}. 
Exotic proton-rich bound systems with \={K}, such as ppK$^-$, pppK$^-$, 
pppnK$^-$ and $^9$B+K$^-$, are expected to be produced 
in (K$^-$, $\pi^-$) reactions \cite{YA}. 
In our previous paper, we investigated kaonic nuclei, which are denoted as \={K} nuclei
 hereafter, $^3$He+K$^-$ 
and $^8$Be+K$^-$, with a simple version of Antisymmetrized Molecular 
Dynamics (AMD) \cite{AMDK}. 
Although our results are similar to those obtained in Ref. \cite{Akaishi-Yamazaki},
 a strange property appeared in  $^8$Be+K$^-$. 
It is an isovector deformation, which means that 
the proton distribution differs from the neutron one in spite of $N=Z$.
Thus, \={K} nuclei seem to provide interesting phenomena 
and stimulate further studies. 

Apparently, these interesting properties of \={K} 
nuclei can be attributed to the \={K}N interaction. 
Especially, the $I=0$ \={K}N interaction plays an essential role. 
According to a precise experiment \cite{Iwasaki:KHatom}, 
the 1s atomic state of a kaonic hydrogen (i.e. proton+K$^-$) 
is shown to be shifted upward. This upward shift, which is consistent with 
the low-energy scattering data of \={K}N \cite{Martin}, 
suggests that the \={K}N interaction is strongly attractive, 
so that the system of a K$^-$ and a proton has 
a nuclear bound state which corresponds to the $I=0$ $\Lambda(1405)$ hyperon 
resonance lying at 27 MeV below the K$^-$p threshold. 
In a boson exchange potential model, the J$\ddot{\rm u}$lich group \cite{Julich} 
showed that all of the $\omega$, $\rho$ and $\sigma$ mesons 
work coherently to give a strong attraction between a \={K} and an N which 
accommodates a K$^-$-p bound state, identified as $\Lambda(1405)$. 
Studies based on chiral SU(3) \cite{Weise} also show that 
the $I=0$ \={K}N interaction is attractive enough to form a $\Lambda(1405)$.
In this paper we employ a phenomenological \={K}N interaction \cite{Akaishi-Yamazaki}, 
whose $t$-matrix in the \={K}N channel is similar to that led by chiral SU(3) theory. 

Since the $I=0$ \={K}N interaction is much more attractive than the $I=1$ one, 
a K$^-$ meson attracts protons rather than neutrons, causing 
an isovector deformation. 
Thus, the $I=0$ \={K}N interaction is essential for studying \={K} nuclei.

In this paper we present systematic studies of \={K} nuclei with AMD.
Since AMD treats a system in a fully microscopic way 
without any assumption concerning the structure of the system, 
it is suitable for studying \={K} nuclei, whose 
structures might be exotic. 
The simple version of AMD \cite{AMDK}, however, has 
a technical problem in treating the 
$I=0$ \={K}N interaction, which dominates \={K} nuclear systems: 
it cannot include the degree of freedom of \={K}$^0$, and fails 
to describe the $I=0$ \={K}N state. Therefore, in our previous paper 
we dealt with the $I=0$ \={K}N interaction by incorporating 
its \={K}$^0$n part effectively into the K$^-$p interaction.
Of course, the $I=0$ \={K}N interaction should be treated 
as precisely as possible, because it plays an essential 
role in \={K} nuclei. 
In this paper, 
we improve the framework of AMD so that it can treat the 
$I=0$ \={K}N interaction adequately. 
We introduce the degree of freedom of \=K$^0$ into the model space of AMD  
 (``pK$^-$/n\={K}$^0$ mixing''). 
Since \={K} nuclear states depend largely 
on their isospin due to the strong isospin-dependence of the 
\={K}N interaction, we carry out the isospin projection 
as well as the angular momentum projection (``$J$ \& $T$ projections'') 
of the obtained intrinsic state.

With the new version of AMD, we systematically investigate 
a variety of \={K} nuclei. 
We try to answer the following questions: 
i) What \={K} nuclei are deeply bound with narrow widths? 
ii) Is there any strange structure peculiar to \={K} nuclei? 

This paper is composed as follows: In Section \ref{Formalism}, 
we present the improvements of AMD;  pK$^-$/n\={K}$^0$ mixing, 
$J$ \& $T$ projections, and other formalisms. 
In Section \ref{Tests of our method}, we demonstrate the capability of our new 
framework, and then apply it to various \={K} nuclei (ppnK$^-$, pppK$^-$, pppnK$^-$, 
$^6$BeK$^-$, and $^9$BK$^-$). The results and a discussion 
are given in Section \ref{Results}. We summarize our study in Section 
\ref{Summary}.

\section{Formalism \label{Formalism}}

In the present study, we employ the AMD approach as a means of studying \={K} nuclei.  
It has succeeded in studying the structures of light unstable nuclei \cite{AMD:Enyo}. 
In particular, it is powerful for investigating 
the global properties of many light nuclei systematically. 
We know that various 
few-body methods, such as summarized in \cite{FewBody}, 
can treat few-body \={K} nuclei more accurately than AMD. 
Compared with these usual methods, AMD is more handy and applicable to more 
complex nuclei.  Our aim is a systematic study of a variety of \={K} nuclei.

In the simple version of AMD employed in our previous study, 
we restricted its model space to the proton, neutron 
and K$^-$ meson. Due to the lack of \={K}$^0$ in the model space, 
we could not describe the $I=0$ \={K}N state,  
\begin{eqnarray}
|\bar{K}N(I=0) \rangle = 
\frac{1}{\sqrt{2}} \left( |p K^- \rangle + |n \bar{K}^0 \rangle \right) .
\end{eqnarray}
In other words, we could not precisely treat the coupling of the pK$^-$ pair with 
the n\={K}$^0$ one through the $I=0$ \={K}N interaction 
in the particle basis treatment of AMD. 
In the previous study, we incorporated all contributions from 
the $I=0$ \={K}N interaction into an effective K$^-$p interaction, as follows: 
\begin{eqnarray}
V_{K^-p} = \alpha \; V_{\bar{K}N (I=0)} + \beta \; V_{\bar{K}N (I=1)} \; , 
\end{eqnarray}
where $\alpha$ and $\beta$ are some constants determined by  
counting the number of $I=0$ pairs and $I=1$ ones in 
a given state of total isospin $T$. For example, we set 
$(\alpha, \beta) = (\frac{3}{4}, \frac{1}{4})$ in 
the case of ppnK$^-$ ($T=0$). 
However, we have to check how 
this prescription is reliable for various \={K} nucleus cases.
For this purpose, we introduce the degree of freedom of \={K}$^0$ 
into the AMD framework to treat ``pK$^-$/n\={K}$^0$ mixing'' directly.

\subsection{pK$^-$/n\={K}$^0$ mixing}

First, we explain our idea for the simple case of a \={K} nucleus ppnK$^-$. 
In this \={K} nucleus, the component of pnn\={K}$^0$
is mixed because 
a pair of pK$^-$ is replaced with that of n\={K}$^0$ 
by the $I=0$ \={K}N interaction. Hereafter, we express this state 
as $^3_{\bar{\rm K}}$H. 
An ordinary way to treat it is to perform a coupled channel calculation, 
preparing several Slater determinants for both channels of ppnK$^-$ and pnn\={K}$^0$. 
In the present paper, we deal with such systems where several channels 
are coupled as follows: 
In stead of multi Slater determinants, we employ a single Slater determinant 
with {\it charge-mixed} single particle wave functions, 
i.e. 
\begin{eqnarray}
| N_i \rangle & = & x_i | p \rangle + y_i | n \rangle, \\
| K  \rangle & = & z   | K^- \rangle + w | \bar{K}^0 \rangle,
\end{eqnarray}
where $|N_i\rangle$ and $| K \rangle$ indicate 
a single nucleon wave function and a \={K} meson wave function, respectively.
$|N_i\rangle$ can describe the state where a proton and a neutron 
are mixed, and also $|K\rangle$ can describe the state where K$^-$ and 
\={K}$^0$ are mixed. With these wave functions we describe $^3_{\bar{\rm K}}$H 
as $|\det[N_1 N_2 N_3] K \rangle$. This state contains 
the component of $| pnn\bar{K}^0 \rangle$ as well as that of $| ppnK^- \rangle$. 
In this method, since each nucleon has a chance to be a proton or a neutron, 
the important channel is automatically chosen in the process of the energy variation. 
In addition, we expect that 
$|\det[N_1 N_2 N_3] K \rangle$ can 
represent a state in which the contribution of several configurations 
is coherently additive: for example, if two configurations such as 
$|ppnK^-\rangle$ and  $|pnn\bar{K}^0\rangle$ work 
coherently, such state is represented as $| p \, (p+n) \, n \, (K^-+\bar{K}^0)\rangle$. 

However, we remark one point: $|\det[N_1 N_2 N_3] K \rangle$ is likely to 
have incorrect components, for example pppK$^-$, ppn\={K}$^0$ and so on, 
which should not couple with $^3_{\bar{\rm K}}$H. 
To avoid the mixing of such incorrect 
components, we project it onto a state whose isospin-z component 
$T_z$ is equal to that of $^3_{\bar{\rm K}}$H.

Now, we show the details of our wave function based on the 
concept of pK$^-$/n\={K}$^0$ mixing. 
Our nucleon wave function, $|\varphi_i \rangle$, 
is represented by the  superposition of several Gaussian wave packets \cite{AMD:Dote}, 
\begin{eqnarray} 
&  |\varphi_i \rangle&   = 
\sum_{\alpha=1}^{N_n} C^i_\alpha 
\exp \left[-\nu \left(\mbox{\boldmath $r$}-\frac{\mbox{\boldmath $Z$}^i_\alpha}{\sqrt{\nu}} 
\right)^2 \right]\; 
|\sigma_i \rangle | \tau^i_\alpha \rangle . \label{eq:N}
\end{eqnarray}
Namely, the $i$-th nucleon is described by the superposition of 
$N_n$ Gaussian wave packets whose centers are $\{ \mbox{\boldmath $Z$}^i_\alpha\}$.
$|\sigma_i \rangle$ means a spin wave function, and is $|\uparrow\rangle$ 
or $|\downarrow\rangle$. $| \tau^i_\alpha \rangle$ means an isospin wave function, 
and has the following form: 
\begin{eqnarray} 
| \tau^i_\alpha \rangle = 
\left( \frac{1}{2}+\gamma^i_\alpha \right)| p \rangle + 
\left( \frac{1}{2}-\gamma^i_\alpha \right)| n \rangle, 
\end{eqnarray} 
where $\gamma^i_\alpha$ is a variational parameter. 
In the usual AMD the isospin of each nucleon does not change, i.e. 
in the process of energy-variation 
the protons remain as protons and the neutrons as neutrons. 
However, in the present paper we make the isospins of all nucleons 
changeable so that we can treat pK$^-$/n\={K}$^0$ mixing. 
In the same way, 
a \={K} meson wave function, $|\varphi_{K} \rangle$, has 
the form 
\begin{eqnarray} 
&  |\varphi_{K} \rangle&   = 
\sum_{\alpha=1}^{N_K} C^{K}_\alpha 
\exp \left[-\nu \left(\mbox{\boldmath $r$}-\frac{\mbox{\boldmath $Z$}^{K}_\alpha}{\sqrt{\nu}} 
\right)^2 \right]\; 
| \tau^{K}_\alpha \rangle . \label{eq:K}
\end{eqnarray}
Here, the isospin wave function of a \={K}, $| \tau^{K}_\alpha \rangle$,
as well as that of a nucleon, is changeable,  
\begin{eqnarray} 
| \tau^{K}_\alpha \rangle = 
\left( \frac{1}{2}+\gamma^{K}_\alpha \right) | \bar{K}^0 \rangle + 
\left( \frac{1}{2}-\gamma^{K}_\alpha \right) |K^-  \rangle .
\end{eqnarray}

Because a nucleon is a fermion, we antisymmetrize the wave function of 
the nucleon's system,  
$|\Phi_{N} \rangle =  \det[|\varphi_i(j)\rangle]$ . 
Then, the \={K} meson wave function is combined to it,   
$|\Phi  \rangle = |\Phi_{N} \rangle \otimes 
|\varphi_{K} \rangle$.
Moreover, we project the total wave function onto the eigen-state of parity, 
\begin{eqnarray}
|\Phi^\pm\rangle & = & \frac{1}{\sqrt{2}}[ \; |\Phi \rangle\pm {\cal P}|\Phi \rangle \; ]. 
\label{eq:pty}
\end{eqnarray}

If we perform an energy-variation with a trial wave function (\ref{eq:pty}), 
it is likely that the z-component of the isospin ($T_z$) of the 
total system is different from 
that of a \={K} nucleus that we try to calculate originally. 
To avoid any mixing of components having an incorrect $T_z$, 
we project the total system 
onto an eigen-state of $T_z$ before the energy-variation,   
\begin{eqnarray}
| \hat{P}_M \Phi^\pm \rangle = 
\int d \theta \; \exp[-i \theta (\hat{T}_z - M)] | \Phi^\pm \rangle .
\end{eqnarray}
Thus, we can obtain a wave function containing only the components of $T_z=M$. 
We utilize $| \hat{P}_M \Phi^\pm \rangle$ as a trial wave function.

Our wave function includes complex number parameters 
$\{X^i_\alpha\} \equiv \{C^i_\alpha, \mbox{\boldmath $Z$}^i_\alpha, 
\gamma^i_\alpha \; ; \;$ $C^K_\alpha, \mbox{\boldmath $Z$}^K_\alpha\, 
\gamma^K_\alpha \}$ 
and a real number parameter ($\nu$). These are determined by the frictional 
cooling equation, as mentioned in \S\S \ref{cooling eq.}.

In the present study, we use a common width parameter ($\nu$) of a Gaussian wave 
packet for a nucleon and for a \={K} meson so as to simplify our calculation.
However, it seems natural that the spreading width of a nucleon is different from that of 
a \={K} meson. We take this point into account by using different numbers of 
Gaussian wave packets for a nucleon and a \={K} meson, i.e. $N_n$ in Eq. (\ref{eq:N}) 
is not equal to $N_K$ in Eq. (\ref{eq:K}).

\subsection{J \& T projections}

The angular momentum projection ($J$ projection) is necessary 
to study  \={K} nuclei as well as that of usual nuclei. 
In addition, the isospin projection ($T$ projection) also seems to 
be important because the \={K}N interaction has strong isospin-dependence.
Therefore, we perform angular-momentum and isospin projections simultaneously.
$J$ projection is done numerically by rotating the system in space,  
as has often been done. $T$ projection is performed in quite the same way, 
but by rotating in isospin space. 
Our $J$ \& $T$ projections are as follows: 
\begin{eqnarray}
| \hat{P}^J_{MK} \; \hat{P}^T_{T_zT_z'} \Phi^\pm \rangle & = &  
\int d\Omega_{ang.} D^{J*}_{MK} (\Omega_{ang.}) \hat{R}_{ang.} (\Omega_{ang.}) \nonumber \\
& & \hspace{-2.0cm} 
\times \int d\Omega_{iso.} D^{T*}_{T_zT_z'} (\Omega_{iso.}) \hat{R}_{iso.} (\Omega_{iso.}) \; 
| \Phi^\pm \rangle , 
\end{eqnarray}
where $| \Phi^\pm \rangle$ is the intrinsic wave function, which is already determined 
by the frictional cooling equation, as shown in \S\S \ref{cooling eq.}. 
We calculate various expectation values with 
$| \hat{P}^J_{MK} \; \hat{P}^T_{T_zT_z'} \Phi^\pm \rangle$.

\subsection{Hamiltonian}

Our Hamiltonian in AMD calculations, 
\begin{eqnarray}
\hat{H} & = & \hat{T} + \hat{V}_{NN}  + \hat{V}_{C} 
+ \hat{V}_{KN} - \hat{T}_G,
\end{eqnarray}
is composed of the kinetic energy $\hat{T}$, 
the effective NN potential $\hat{V}_{NN}$, the Coulomb force 
$\hat{V}_{C}$, and the effective \={K}N potential $\hat{V}_{KN}$. 
The center-of-mass motion energy, $\hat{T}_G$, is subtracted. 
In the kinetic energy and the center-of-mass motion energy, 
we treat the mass difference between a nucleon and a \={K} meson correctly. 
For example, the kinetic energy is 
\begin{eqnarray}
\hat{T} = \sum_{i=1}^{A} \frac{\hat{\mbox{\boldmath $p$}}_i^2}{2 m_{N}} + 
\frac{\hat{\mbox{\boldmath $p$}}_K^2}{2 m_{K}} , 
\end{eqnarray}
where $m_{N}$ and $m_{K}$ indicate the mass of a 
nucleon and that of K$^-$, respectively. 
The Coulomb force is represented by the superposition 
of seven-range Gaussians \cite{AMD:OnoDron}. 

In the study of \={K} nuclei, we do not use existing effective interactions 
which may be justified  for studying phenomena around the normal density. 
Since the system is likely to become extremely dense due to 
the strong \={K} attraction, we employ the $g$-matrix method \cite{Akaishi-Yamazaki}. 
We adopt the Tamagaki potential (OPEG) \cite{Tamagaki} as a bare NN interaction, and 
 the Akaishi-Yamazaki \={K}N potential\cite{Akaishi-Yamazaki} as a bare \={K}N interaction. 
Because the Tamagaki potential can 
reproduce NN phase shifts up to 660 MeV in the lab. \cite{Tamagaki}, 
we expect that it can be applied to such extremely dense states. 
The effective NN/\={K}N interactions 
constructed from the bare ones are represented by the following ten-range Gaussians: 
\begin{eqnarray}
V_{NN}^{\rm X} (r) & = & \sum_{a=1}^{10} V_{NN,a}^{\rm X} \exp[-(r/r_a)^2] \label{Eq:VNN} \\
V_{\bar{K}N}^{I} (r) & = & \sum_{a=1}^{10} V_{\bar{K}N,a}^{I} \exp[-(r/r_a)^2] \label{Eq:VKN} 
\end{eqnarray}
``X'' in Eq. (\ref{Eq:VNN}) is $^1E, \; ^3E, \; ^1O \; {\rm or} \; ^3O$, 
and ``$I$'' in Eq. (\ref{Eq:VKN}) is 0 or 1.
We use these $V_{NN}^{\rm X} (r)$ and $V_{\bar{K}N}^I (r)$ as 
effective NN/\={K}N interactions in our AMD calculation. 

Our procedure is as follows: 
1) For a given density and starting energy of K$^-$, 
we construct a $g$-matrix.
2) Using the $g$-matrix we carry out the AMD calculation. 
3) After the AMD calculation, we check whether or not the obtained density 
and binding energy of K$^-$ are consistent with those of 
the $g$-matrix used in the calculation.
4) If no consistency is accomplished, we guess and impose a new 
density and a new starting energy of K$^-$ for the $g$-matrix calculation 
by referring to the results obtained so far and return to 1).
We repeat this cycle until obtaining a consistent result.

\subsection{Frictional cooling equation with constraint \label{cooling eq.}}

Our wave function contains complex variational parameters,  
$\{X^i_\alpha\}=\{C^i_\alpha, \mbox{\boldmath $Z$}^i_\alpha, 
\gamma^i_\alpha \; ; \;$ $C^K_\alpha, \mbox{\boldmath $Z$}^K_\alpha\, \gamma^K_\alpha \}$. 
They are determined by the energy-variation. In our study, 
we employ the frictional cooling method as a means of energy-variation, 
\begin{eqnarray}
\dot{X}^i_\alpha & = & (\lambda+i\mu)\frac {1}{i \hbar} 
\left[ \frac{\partial {\cal H}}{\partial X^{i*}_\alpha}  
+ \eta \frac{\partial {\cal W}}{\partial X^{i*}_\alpha} \right] 
\hspace{0.2cm} {\rm and} 
\hspace{0.2cm} {\rm C.C.}  \label{eq:cool}
\end{eqnarray}
Here, ${\cal H}$ is the expectation value of the Hamiltonian and 
${\cal W}$ is a constraint condition. $\eta$ is a Lagrange multiplier,  
which is determined by $d{\cal W}/dt=0$. 
It is easily proved that, 
if we assume $\mu < 0$ in Eq. (\ref{eq:cool}) and all of the parameters are developed 
with time according to Eq. (\ref{eq:cool}), the energy of the system decreases  
while satisfying the constraint condition ${\cal W}=0$. 
If we use the superposition of several Gaussian wave packets to represent a 
nucleon and a \={K} meson wave function, 
we need a constraint condition in order to fix 
the center of mass of the total system to the origin, and then 
${\cal W}$ is expressed as follows:
\begin{eqnarray}
{\cal W} & = &  \langle \hat{\mbox{\boldmath $R$}}_G \rangle ^2 + 
\langle \hat{\mbox{\boldmath $P$}}_G \rangle ^2 , \\
& & \hat{\mbox{\boldmath $R$}}_G =  \frac{\sum_{i=1}^A m_N \hat{\mbox{\boldmath $r$}}_i 
+ m_K \hat{\mbox{\boldmath $r$}}_K}{A m_N + m_K} \; , \\ 
& & \hat{\mbox{\boldmath $P$}}_G = \sum_{i=1}^A \hat{\mbox{\boldmath $p$}}_i + 
\hat{\mbox{\boldmath $p$}}_K .
\end{eqnarray}


\section{Tests of our method \label{Tests of our method}}

Before applying our method to studies of various \={K} nuclei, 
we investigate the basic properties of our method. 

\subsection{Dependence on the number of wave packets}

\begin{table}[t]
\caption{Results of ppnK$^-$ in various $N_n$ and $N_K$. 
B.E.: total binding energy. $\rho(0)$: central density. $R_{rms}$, $R_{rms}^p$, $R_{rms}^n$: 
root-mean-square radii of matter, proton and neutron, respectively. 
($\beta$, $\gamma$): deformation parameters.}
\begin{ruledtabular}
\begin{tabular}{cc|ccccccc}
  $N_n$      & $N_K$      &  B.E. & $\rho(0)$ & $R_{rms}$  & $R_{rms}^p$  & $R_{rms}^n$ 
&  $\beta$  & $\gamma$ \\
             &            & [MeV] & [fm$^{-3}$] & [fm] &  [fm] &  [fm] &           & [deg.] \\
\hline \hline
       2     &    5       & 112.7 & 1.39  & 0.72   & 0.70   & 0.75   & 0.19      & 0.0 \\
       3     &    5       & 113.8 & 1.40  & 0.73   & 0.70   & 0.77   & 0.14      & 49.0 \\
       4     &    5       & 117.0 & 1.41  & 0.72   & 0.69   & 0.78   & 0.02      & 49.4 \\
\hline
       2     &    8       & 113.5 & 1.37  & 0.72   & 0.71   & 0.75   & 0.19      & 0.0 \\
       2     &    10      & 113.7 & 1.37  & 0.72   & 0.71   & 0.75   & 0.18      & 0.0 \\
\hline
       3     &    10      & 115.1 & 1.41  & 0.72   & 0.70   & 0.77   & 0.14      & 51.7 \\
       4     &    10      & 116.9 & 1.49  & 0.71   & 0.68   & 0.77   & 0.00      & 54.4 \\
\end{tabular}
\end{ruledtabular}
\end{table}

As shown in Eq. (\ref{eq:N}) and Eq. (\ref{eq:K}), we represent a single 
nucleon wave function and a \={K} meson  wave function 
with $N_n$ and $N_K$ Gaussian wave packets, respectively. 
We investigate how much the solution depends on $N_n$ 
and $N_K$. First, we perform a test in the case of ppnK$^-$ 
without pK$^-$/n\={K}$^0$ mixing for simplicity. 
Table I shows the results of ppnK$^-$ for various $N_n$ 
and $N_K$. From this table, we find that the total binding energy and 
the central density are almost converged up to $N_n=4$ and $N_K=10$. 
However, we notice that the shape of the system, represented by 
the deformation parameters ($\beta$, $\gamma$), is strongly dependent of $N_n$. 
This phenomenon can be understood as follows.  
As mentioned in our previous study \cite{AMDK}, 
the protons distribute compactly near a K$^-$ so as to 
decrease their total energy by the strongly attractive K$^-$p
interaction. On the other hand, the neutron is widely spread and
its total energy decreases by reducing its kinetic energy.
Therefore, the protons stay compactly inside the system, 
while the neutron remains widely outside of the system. 
Thus, the neutron contributes to the shape of the total system. 
In the case of $N_n=2$, 
since the neutron is represented by two Gaussian wave packets, 
it can spread only linearly. Thus, the total system deforms prolately.
In the case of $N_n=3$,
it can spread with a triangular shape. Thus, the total system
deforms oblately. 
In the case of $N_n=4$, 
it can spread with a tetrahedron shape.
The total system is therefore spherical.
Thus, the shape of the total system changes as $N_n$ is varied.

Such a dependence of the shape on $N_n$ seems to be peculiar 
to ppnK$^-$ where the neutron number is equal to 1 and the proton number 
is 2. In addition,  
the total binding energy and the central density do not so strongly 
depend on $N_n$ and $N_K$. Therefore, taking the cost-performance of 
calculations into account, we adopt the model space of 
$N_n=2$ and $N_K=5$ in our calculations.

\subsection{Solution of ppnK$^-$}

\begin{table}[b]
\caption{Quantum numbers before and after projection. 
$J^2_T$: $\langle \hat{\mbox{\boldmath $J$}}^2 \rangle$ of total system. 
$J^2_N$, $L^2_N$ and $S^2_N$: $\langle \hat{\mbox{\boldmath $J$}}^2 \rangle$, 
$\langle \hat{\mbox{\boldmath $L$}}^2 \rangle$, and 
$\langle \hat{\mbox{\boldmath $S$}}^2 \rangle$ of nucleon system. 
$L^2_K$: $\langle \hat{\mbox{\boldmath $L$}}^2 \rangle$ of a \={K} meson.
$T^2$ and $T_z$: $\langle \hat{\mbox{\boldmath $T$}}^2 \rangle$ and 
$\langle \hat{T}^2_z \rangle$ 
of total system.}
\begin{ruledtabular}
\begin{tabular}{l|ccccc|cc}
        & $J^2_T$ & $J^2_N$ & $L^2_N$ & $S^2_N$ & $L^2_K$ & $T^2$ & $T_z$ \\  
\hline
After   & 0.75    & 0.78    & 0.03    & 0.75    & 0.03    & 0.00  & 0.00  \\
Before  & 1.36    & 1.22    & 0.44    & 0.78    & 0.14    & 0.02  & 0.00  \\
\end{tabular}
\end{ruledtabular}
\end{table}

We now check whether 
our new framework, pK$^-$/n\={K}$^0$ mixing and $J$ \& $T$ projections, works 
correctly or not. We perform a test on a system of ppnK$^-$. 

First, we investigate the property of $J$ \& $T$ projections. 
Although only the $J$ projection has often been carried out  
in the study of light unstable nuclei \cite{AMD:Enyo,AMD:Dote}, 
the present study for the first time makes the $T$ projection. 
In Table II, we show various quantum numbers of the wave function 
before projection 
($| \hat{P}_M \Phi^\pm \rangle $) and that after projection 
($| \hat{P}^J_{MK} \; \hat{P}^T_{T_zT_z'} \Phi^\pm \rangle$). 
Apparently, the ground state of ppnK$^-$ seems to have quantum numbers of 
$J^\pi=\frac{1}{2}^+$ and $T=0$. 
We performed $J$ \& $T$ projections so that 
the total system had such quantum numbers. Table II shows that  
$\langle \hat{\mbox{\boldmath $J$}}^2 \rangle=0.75$ and 
$\langle \hat{\mbox{\boldmath $T$}}^2 \rangle=0.00$, 
which agree with $J(J+1)=\frac{1}{2} \cdot \frac{3}{2}$ and 
$T(T+1)=0 \cdot 1$, respectively. Therefore, it is found that our $J$ \& $T$ projections 
work well. 

Next, in Table III various quantities obtained in the present calculation (Present) 
are compared with our previous result of a simple version of AMD (simple AMD) \cite{AMDK}
and the result of a BHF calculation (BHF) \cite{Akaishi-Yamazaki}. 
This table shows that the present result is almost identical to others. 
Since the isospin-z component of each particle is 
changeable in the present framework, 
we investigated each particle-number. 
Although we calculated ppnK$^-$, 
the numbers of protons and neutrons after the calculation 
are both equal to 1.5, while 
those of K$^-$ and \={K}$^0$ are both 0.5. 
This means that ppnK$^-$ and pnn\={K}$^0$ are mixed with  
a ratio of 1:1 as the result of pK$^-$/n\={K}$^0$ coupling through 
the $I=0$ \={K}N interaction. 

Here, we remark on the components of the \={K}N interaction. 
We can separate it into three parts in the particle base: 
i) $V_{nK^-}$ and $V_{p\bar{K}^0}$, 
ii) $V_{pK^-}$ and $V_{n\bar{K}^0}$, and iii) $V_{pK^-,n\bar{K}^0}$. 
The interactions i) and ii) are working in each channel of 
ppnK$^-$ and pnn\={K}$^0$, and their expectation values are equal to 
$-45$ MeV and $-255$ MeV, respectively. Interaction iii) is related 
to pK$^-$/n\={K}$^0$ mixing through the $I=0$ \={K}N interaction, 
and its expectation value is equal to $-88$ MeV.
Thus, we find that the binding of this system is mainly due to 
the type ii) interaction and is further supported by 
the type iii) interaction, which 
causes coupling between the two channels. 

\begin{table}
\caption{Results of ppnK$^-$.
B.E.: binding energy measured from the threshold of ppn+K$^-$.
$\Gamma$: width decaying to ${\rm \Lambda}\pi$. 
$\rho(0)$: central density. 
$R_{rms}$: root-mean-square radius of nucleon system.} 
\begin{ruledtabular}
\begin{tabular}{l|cccc}
         &  B.E.  &  $\Gamma$ &  $\rho(0)$        &  $R_{rms}$  \\
         &  [MeV] &  [MeV]    &  [fm$^{-3}$]  &  [fm]  \\     
\hline
Present  & 110.3  & 21.2      & 1.50          & 0.72      \\
simple AMD & 105.2  & 23.7      & 1.39          & 0.72      \\
BHF  & 108    & 20        & ---           & 0.97      \\
\end{tabular}
\end{ruledtabular}
\end{table}

\subsection{Interpretation of the density distribution}

\begin{figure}[t]
\begin{minipage}[t]{3cm}
   \includegraphics[width=1.00\textwidth]{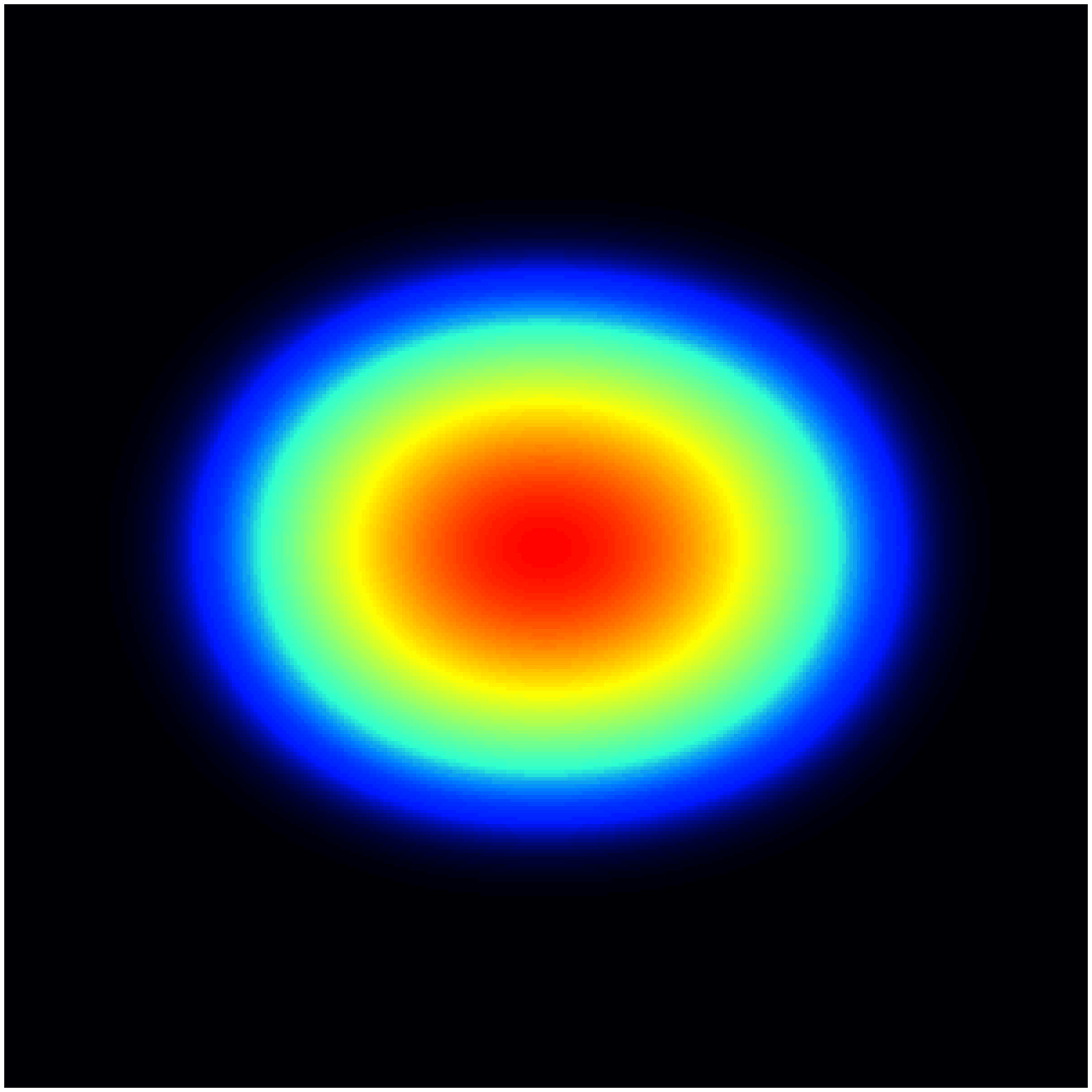}%
\end{minipage}
\hspace{0.3cm}
\begin{minipage}[t]{3cm}
   \includegraphics[width=1.00\textwidth]{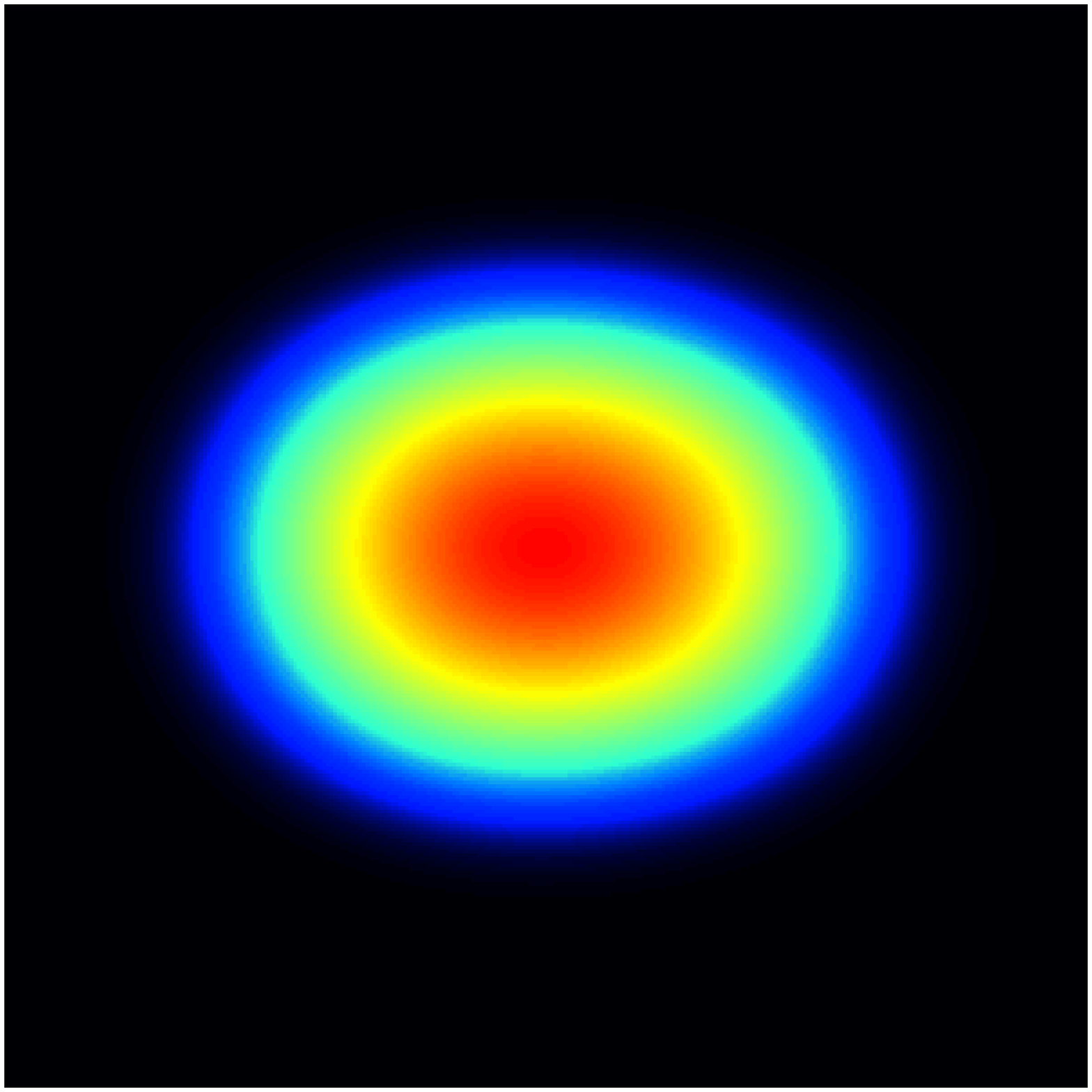}%
\end{minipage}

   \caption{\label{ppnK1}
Proton (left) and neutron (right) distributions of ppnK$^-$ 
obtained in the present calculation. 
The size of each frame is 3fm $\times$ 3fm. 
}
\end{figure}

The influence of pK$^-$/n\={K}$^0$ mixing can clearly be seen in the 
density distribution. Fig. \ref{ppnK1} displays the density distribution 
of protons and neutrons in the ppnK$^-$ system calculated by the 
new framework. We can see that the proton distribution is almost 
the same as the neutron one, contrary to our previous 
result \cite{AMDK}, where protons distribute more compactly than neutrons 
because of the strong attraction between the K$^-$ and the proton. 
We can solve this contradiction by introducing the concept of 
an {\it intrinsic state in isospin space}, as follows: 
the calculated expectation value of $\mbox{\boldmath $T$}^2$ with the state 
$|\hat{P}_{T_z=0} \; \Phi \rangle$, in the case of ppnK$^-$, 
is nearly equal to zero. Therefore, this state is 
the eigen-state of isospin, i.e. $T=0$, and we express it 
as $|^3_{\bar{\rm K}}{\rm H} \; (T=0)\rangle$ hereafter. 
It is easily found that $|^3_{\bar{\rm K}}{\rm H} \; (T=0)\rangle$ is composed of 
two configurations concerning the z-component of isospin: 

\begin{widetext}
\begin{eqnarray}
|^3_{\bar{\rm K}}{\rm H}(T=0)\rangle 
& = & \hat{P}_{T_z=0} \; \left[|\Phi_N\rangle \otimes |\varphi_K\rangle \right] \nonumber \\
& = & \hat{P}_{T_z=0} \; \left[ \; \left(\sum_{m=-\infty}^{+\infty} \hat{P}_{T^N_z=m}
\right)|\Phi_N\rangle 
\otimes \left(\sum_{m=\pm 1/2} \hat{P}_{T^K_z=m} \right)|\varphi_K\rangle \; \right] 
\nonumber\\
& = & \left| \hat{P}_{T^N_z=\frac{1}{2}} \, \Phi_N \right\rangle \otimes 
\left| \hat{P}_{T^K_z=-\frac{1}{2}} \, \varphi_K \right\rangle 
 + \left| \hat{P}_{T^N_z=-\frac{1}{2}} \, \Phi_N \right\rangle \otimes 
\left| \hat{P}_{T^K_z=\frac{1}{2}} \, \varphi_K \right\rangle,
\end{eqnarray}
\end{widetext}
where $\hat{P}_{T^N_z}$ and $\hat{P}_{T^K_z}$ are $T_z$-projection operators 
for the nucleon system and the \={K} meson, respectively. 
According to the values of $T^N_z$ and $T^K_z$, 
the first term indicates ppnK$^-$, while the second term indicates pnn\={K}$^0$. 
Hereafter, we express them as $| ppnK^- \rangle$ and 
$| pnn\bar{K}^0 \rangle$, respectively. 
In addition, 
the overlap between $| pnn\bar{K}^0 \rangle$ 
and $e^{i \pi \hat{T}_y}| ppnK^- \rangle$, calculated numerically, 
is found to be almost one. Since this fact indicates 
$| pnn\bar{K}^0 \rangle \simeq 
e^{i \pi \hat{T}_y}| ppnK^- \rangle$, we can say that 
our wave function satisfies 
\begin{eqnarray}
|^3_{\bar{\rm K}}{\rm H}(T=0)\rangle = 
\sum_{\theta=0,\pi} e^{i \, \theta \, \hat{T}_y} \; | ppnK^- \rangle \nonumber \\
= \sum_{\theta=0,\pi} e^{i \, \theta \, \hat{T}_y} \; | pnn\bar{K}^0 \rangle
.
\end{eqnarray}

Namely, the $| ppnK^- \rangle$ state rotates in isospin space 
so that  
$|^3_{\bar{\rm K}}{\rm H}(T=0)\rangle$, which has the 
good quantum number, $T=0$, is formed. 
Based on the analogy of an ``intrinsic'' state in a deformed nucleus, 
rotating in the space to form the eigen-state of the angular momentum, 
we can regard the $| ppnK^- \rangle$ or 
equivalent $| pnn\bar{K}^0 \rangle$ as 
an {\it intrinsic state in the isospin space} of 
$|^3_{\bar{\rm K}}{\rm H}(T=0)\rangle$. 

\begin{figure}[t]
\begin{minipage}[t]{3cm}
   \includegraphics[width=1.00\textwidth]{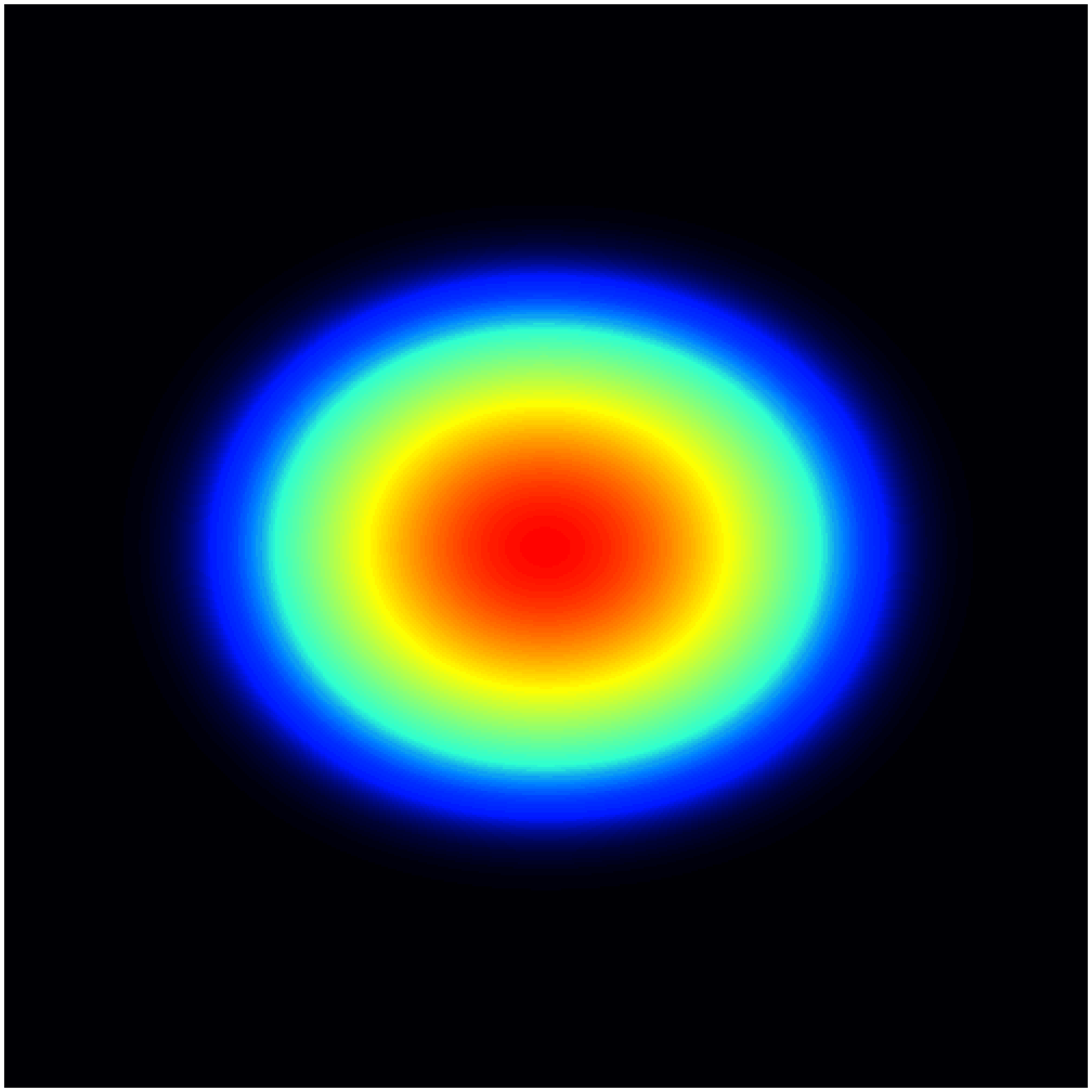}%
\end{minipage}
\hspace{0.3cm}
\begin{minipage}[t]{3cm}
   \includegraphics[width=1.00\textwidth]{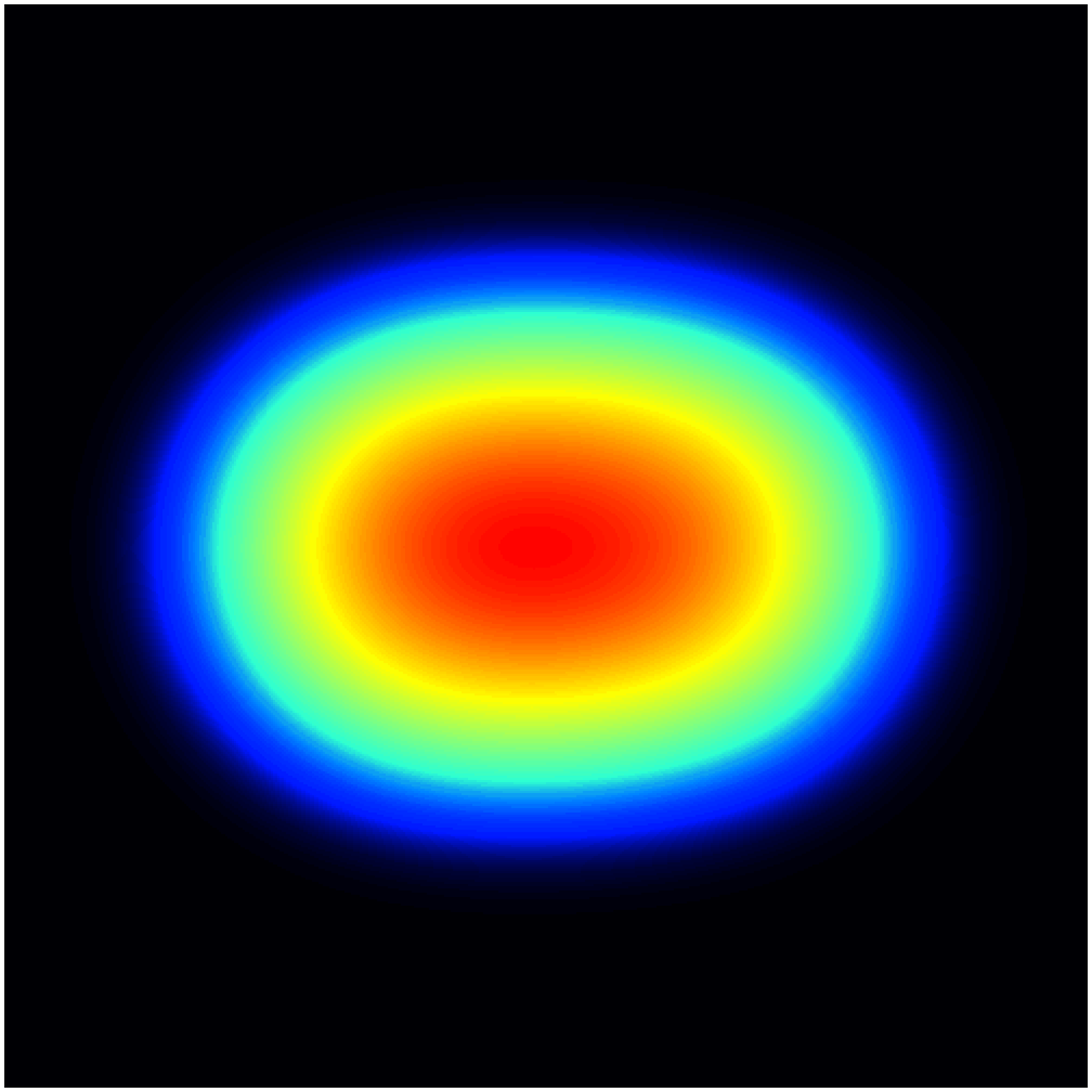}%
\end{minipage}

   \caption{\label{ppnK2}
Proton (left) and neutron (right) distributions of ppnK$^-$. 
See the text for details. 
}
\end{figure}

Fig. \ref{ppnK2} displays the proton and neutron distributions of the 
intrinsic state $| ppnK^- \rangle$. Clearly, the proton 
distribution is more compact than the neutron one; this fact is 
consistent with our previous study. 
Here, we note one point. Investigating the density distribution of 
$| pnn\bar{K}^0 \rangle$, we find that its proton and neutron 
distributions are completely identical with the neutron and proton 
ones in the $| ppnK^- \rangle$, respectively. In other words, 
neutron distribution is more compact than proton one 
in the $| pnn\bar{K}^0 \rangle$. 
After all, since Fig. \ref{ppnK1} is drawn with 
the $|^3_{\bar{\rm K}}{\rm H}(T=0)\rangle$, which includes 
the $| ppnK^- \rangle$ and the $| pnn\bar{K}^0 \rangle$
to form the eigen-state of isospin, the proton distribution is 
quite the same as the neutron one.

\section{Results \label{Results}}

Now that we have confirmed that the new framework works well, 
we proceed to an investigation of 
pppK$^-$, pppnK$^-$, $^6$BeK$^-$ and $^9$BK$^-$ with this 
framework. 
Since the intrinsic state of isospin space is found to be meaningful,  
as mentioned in the previous section, 
we generally denote \={K} nuclei by their intrinsic states without any loss 
of validity; for example, $^3_{\bar{\rm K}}$H is represented by ppnK$^-$. 
In all calculations, a single nucleon and \={K} meson are described 
with two Gaussian wave packets and five, respectively, 
namely $N_n=2$ in Eq. (\ref{eq:N}) and $N_K=5$ in Eq. (\ref{eq:K}).  

\begin{figure*}
   \includegraphics[width=0.8\textwidth]{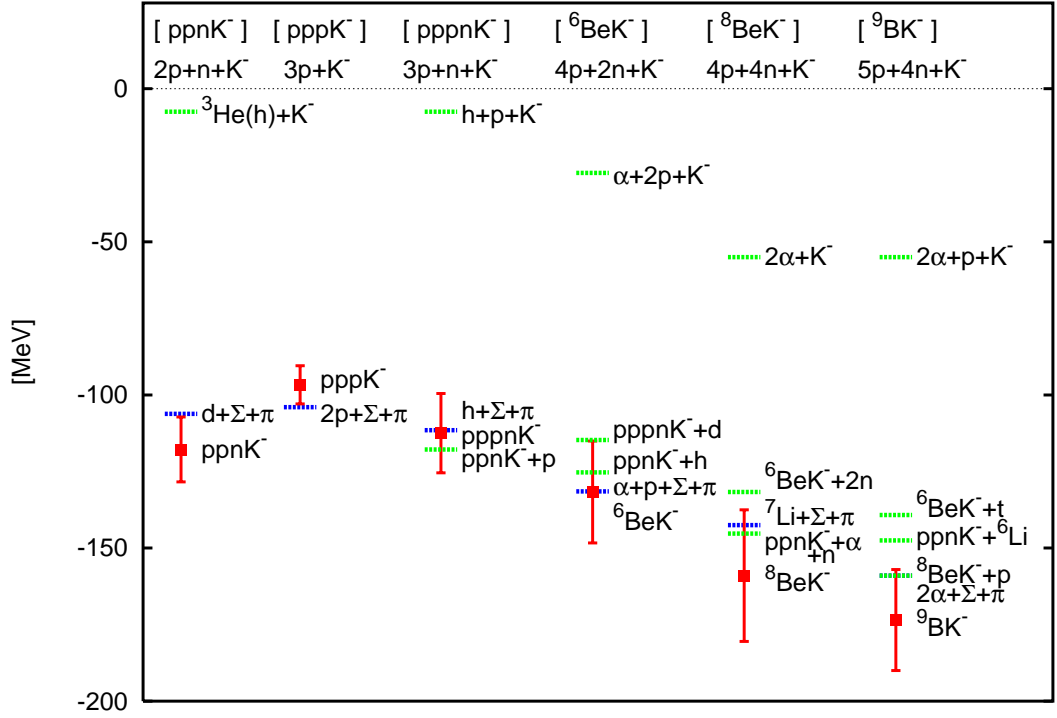}%

   \caption{\label{Energies}
Behavior of the total binding energy and the decay width, 
from ppnK$^-$ to $^9$BK$^-$.
The decay width is to ${\rm \Lambda}\pi$ and ${\rm \Sigma}\pi$ channels. 
The red point and line indicates the binding energy and width of 
the \={K} nucleus.
The blue-dashed line corresponds to the ${\rm \Sigma}\pi$ threshold. 
The thresholds for other decay modes are expressed by 
the green-dashed lines.}
\end{figure*}

\begin{figure*}
\begin{minipage}[t]{3.6cm}
   \includegraphics[width=1.00\textwidth]{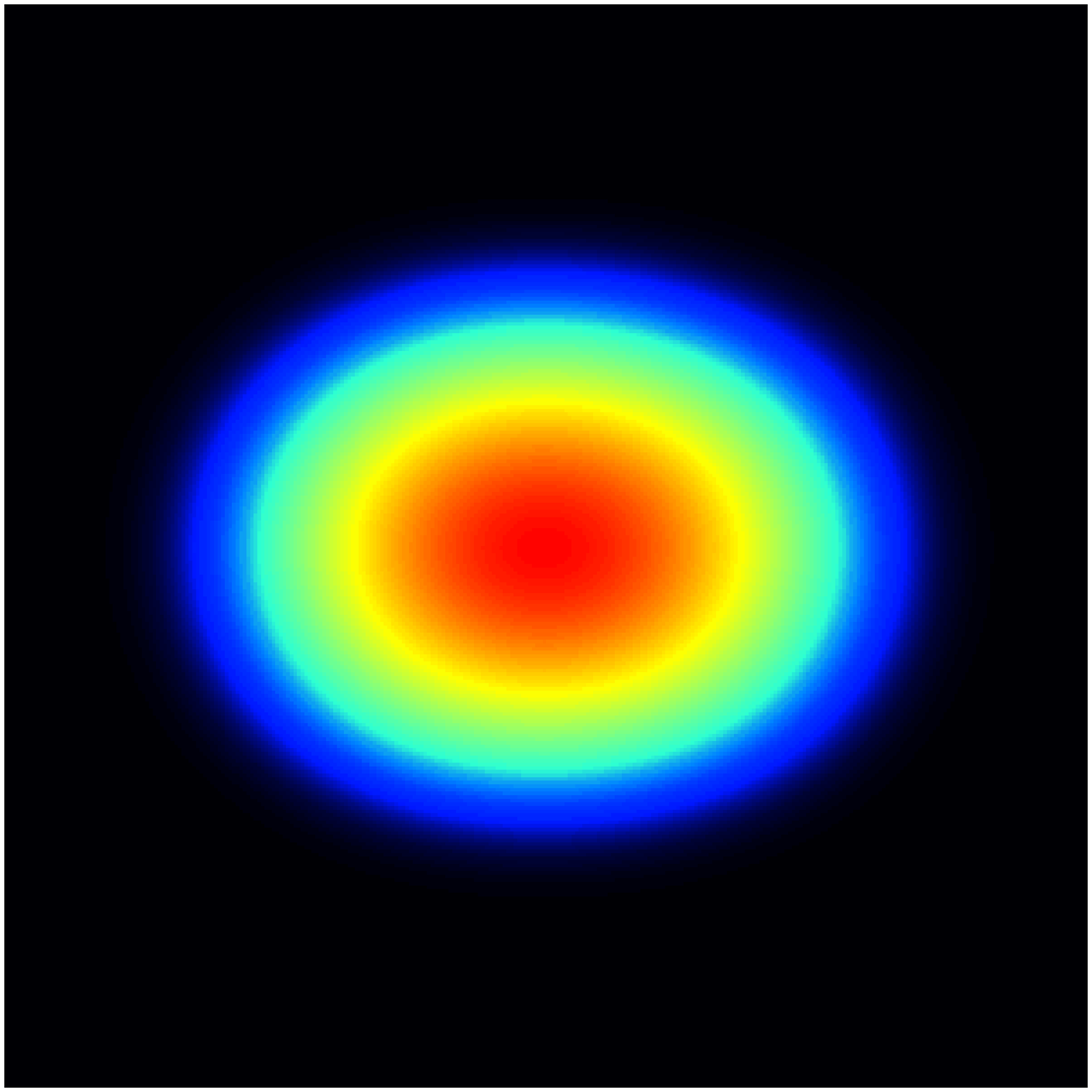}%

ppn{\Km}
\end{minipage}
\hspace{0.3cm}
\begin{minipage}[t]{3.6cm}
   \includegraphics[width=1.00\textwidth]{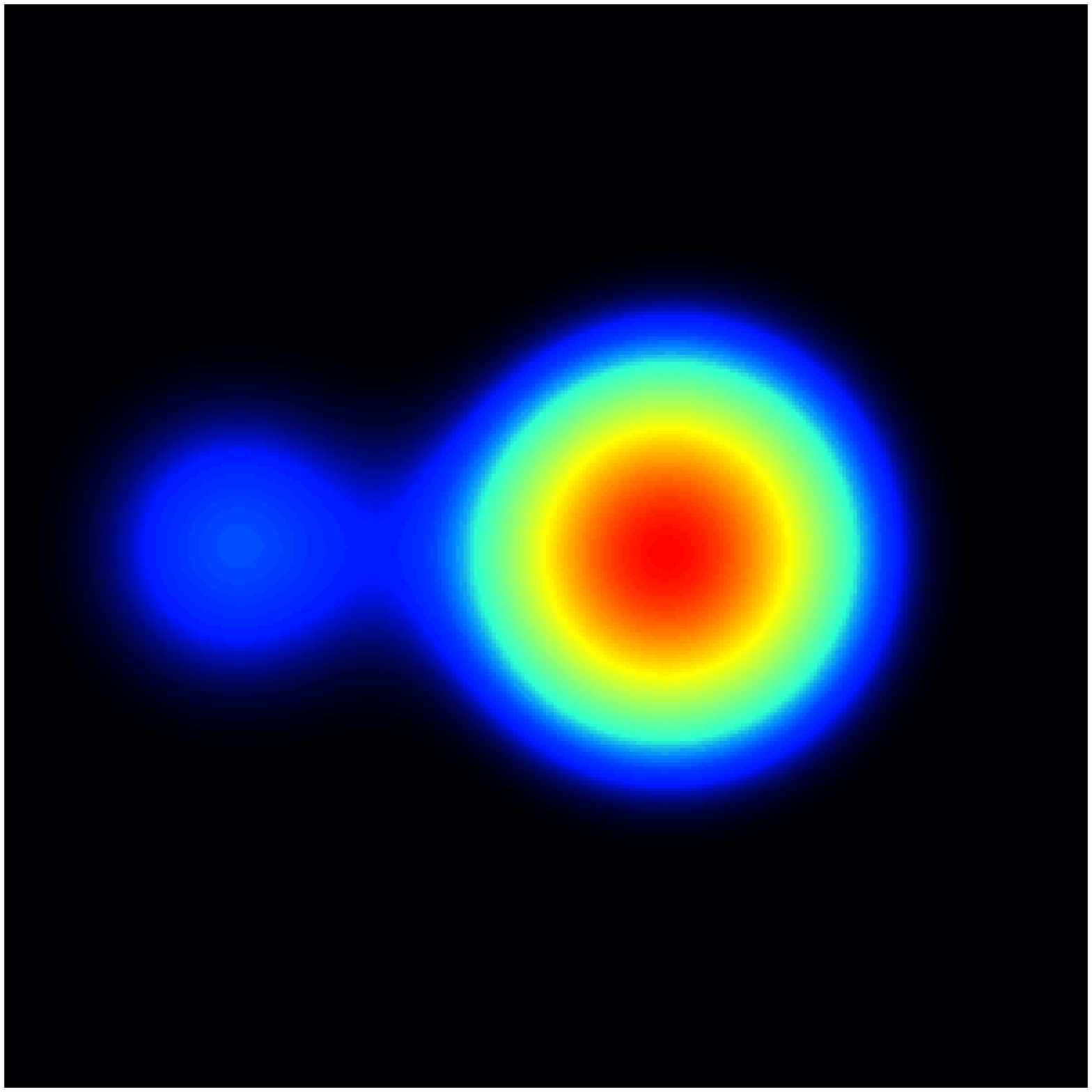}%

ppp{\Km}
\end{minipage}
\hspace{0.3cm}
\begin{minipage}[t]{3.6cm}
   \includegraphics[width=1.00\textwidth]{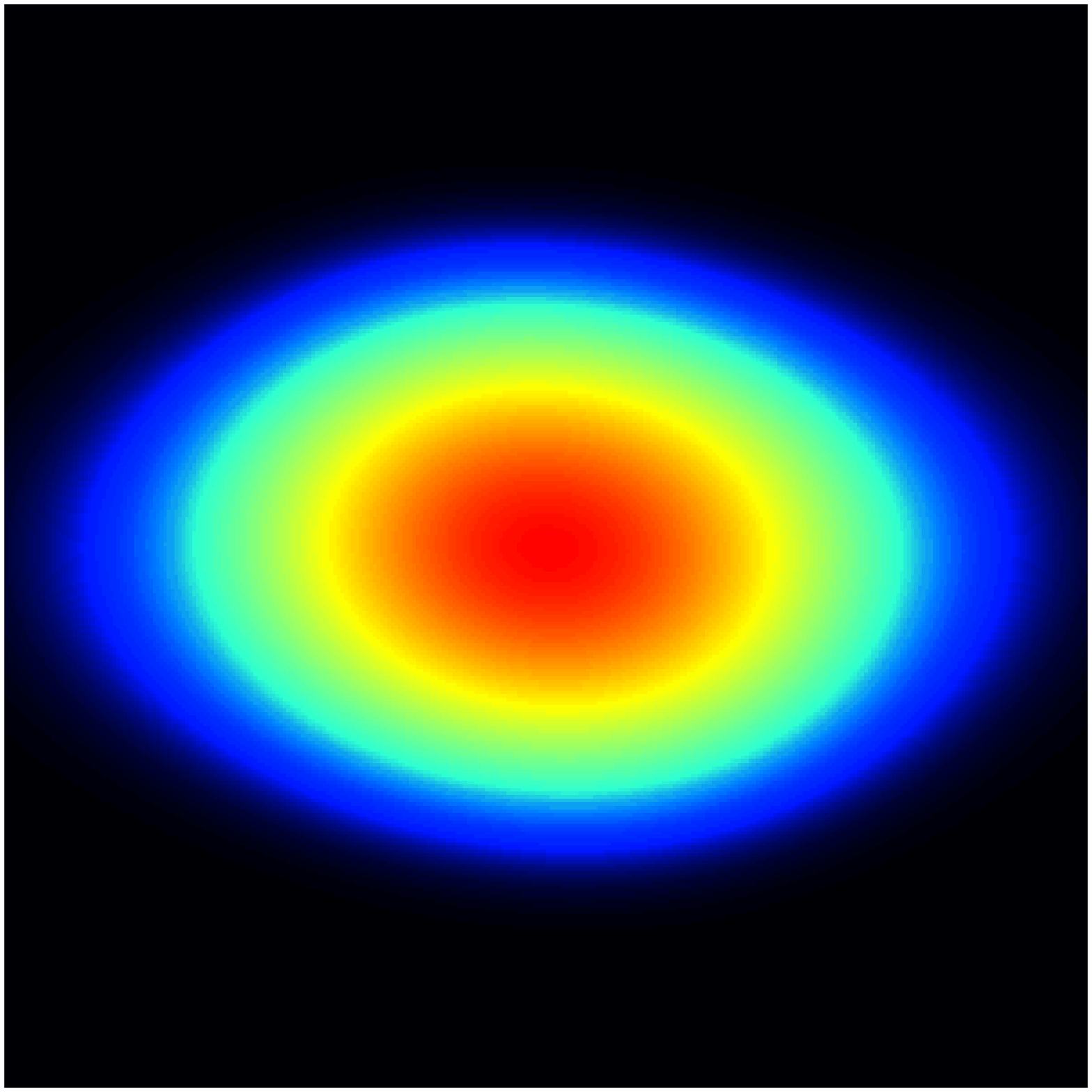}%

pppn{\Km}
\end{minipage}

\vspace{0.5cm}

\begin{minipage}[t]{4.8cm}
   \includegraphics[width=1.00\textwidth]{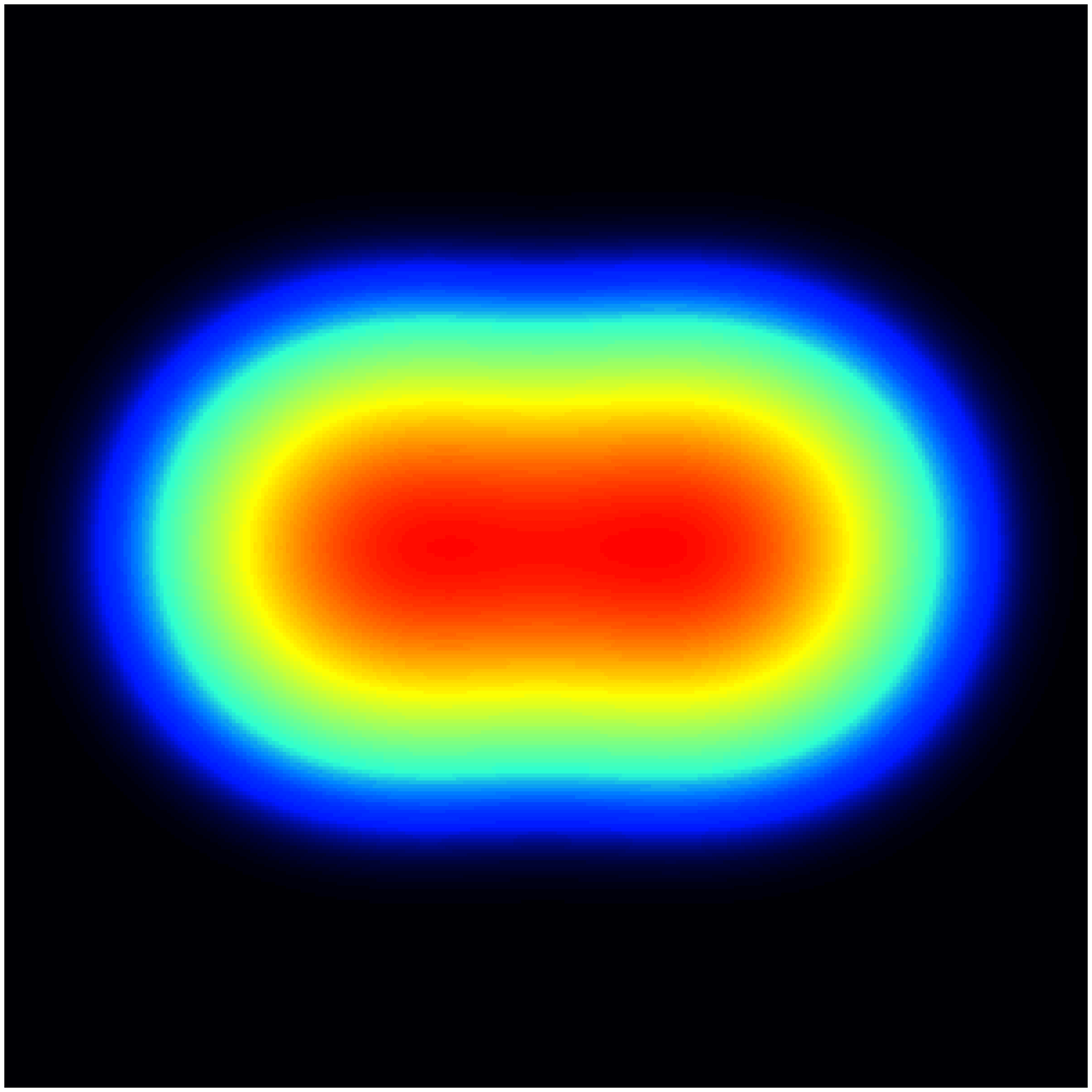}%

$^6$Be{\Km}
\end{minipage}
\hspace{0.1cm}
\begin{minipage}[t]{4.8cm}
   \includegraphics[width=1.00\textwidth]{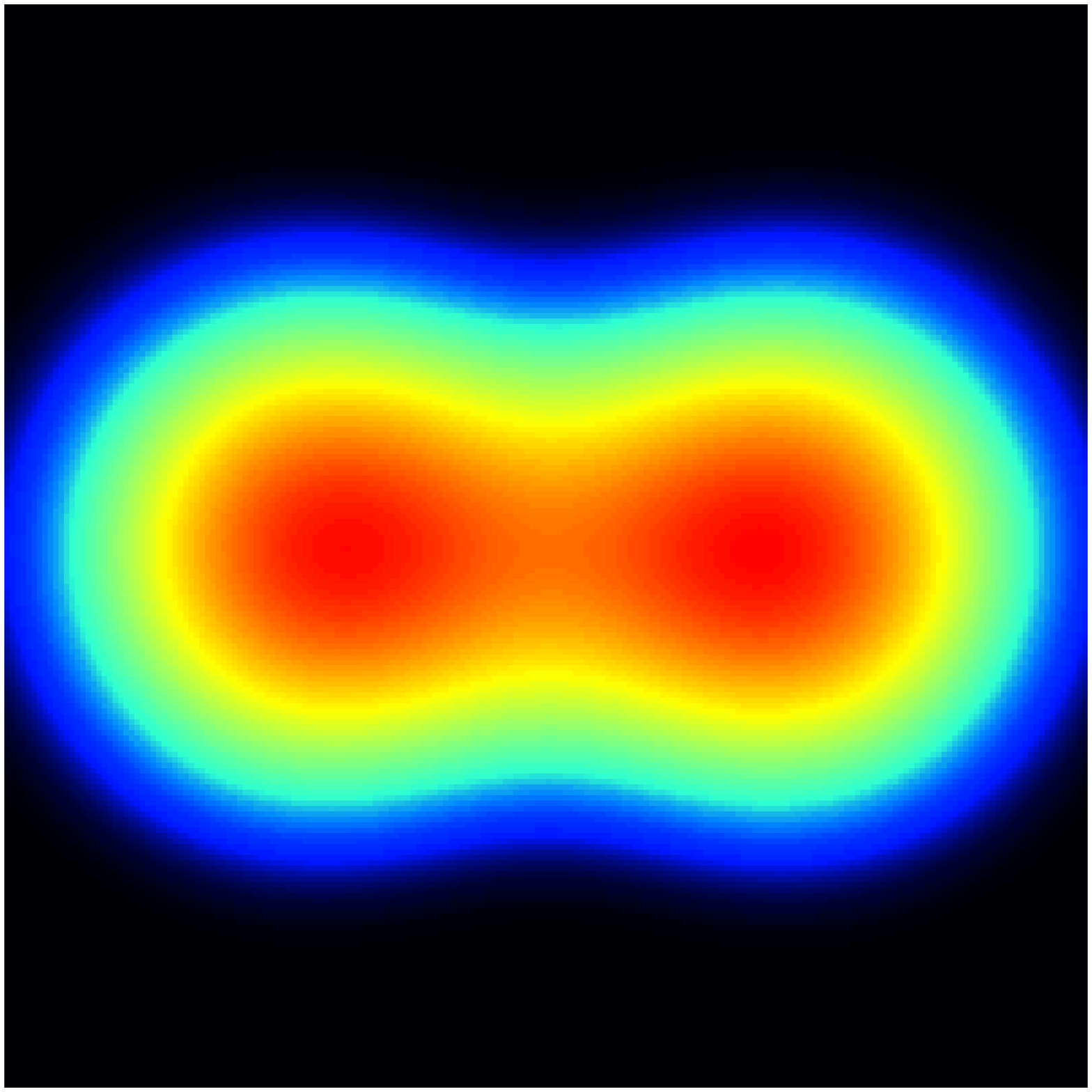}%

$^9$B{\Km}
\end{minipage}

   \caption{\label{Densities}
Density contours of 
the nucleon distributions of various \={K} nuclei. 
ppnK$^-$, pppK$^-$ and pppnK$^-$: 3fm $\times$ 3fm. 
$^6$BeK$^-$ and $^9$BK$^-$: 4fm $\times$ 4fm.}
\end{figure*}

\subsection{Binding energies}

\begin{table}
\caption{Summary of present calculations. 
$J^\pi$ and $T$: spin-parity and isospin of total system.  
E(\={K}): binding energy of K$^-$ meson. 
$\Gamma$: width decaying to ${\rm \Sigma}\pi$ and ${\rm \Lambda}\pi$ channels. 
$\rho(0)$: central density. 
$R_{rms}$: root-mean-square radius of nucleon system. 
($\beta$, $\gamma$): deformation parameters.}
\begin{ruledtabular}
\begin{tabular}{l|cc|cccccc}
              & $J^\pi$ & $T$ & E(\={K})   & $\Gamma$ & $\rho(0)$ & $R_{rms}$ & $\beta$ & $\gamma$  \\
              &         &   & [MeV]  & [MeV]    & [fm$^{-3}$] & [fm]  &         & [deg.] \\
\hline
ppnK$^-$      & $\frac{1}{2}^+$ & 0 & 110.3     & 21.2        & 1.50   & 0.72     & 0.22    & 9.2             \\
pppK$^- $     & $\frac{3}{2}^-$ & 1 &  96.7     & 12.5        & 1.56   & 0.81     & 0.70    & 11.8            \\
pppnK$^-$     & $1^-$ & $\frac{1}{2}$ & 105.0     & 25.9        & 1.29  & 0.97    & 0.54    & 3.8             \\
$^6$BeK$^-$   & $0^+$ & $\frac{1}{2}$ & 104.2     & 33.3        & 0.91  & 1.17    & 0.44    & 0.3             \\
$^9$BK$^-$   & $\frac{3}{2}^-$ & 0  & 118.5     & 33.0        & 0.71  & 1.45    & 0.46    & 20.8             \\
\end{tabular}
\end{ruledtabular}
\end{table}

\begin{table}[b]
\caption{Binding energy and number of strongly interacting nucleons near the \={K} meson.}
\begin{ruledtabular}
\begin{tabular}{l|ccccc}
         & ppnK$^-$   &  pppK$^-$ & pppnK$^-$ & $^6$BeK$^-$ & $^9$BK$^-$ \\
\hline
B.E.     & 110.3      &  96.7     & 105.0     & 104.2       & 118.5      \\
Nucleon  & 1.67       &  1.14     & 1.78      & 2.55        & 2.53       \\
\end{tabular}
\end{ruledtabular}
\end{table}

The results are summarized in Table IV. 
We determine $J^\pi$ and $T$ by assuming that 
nucleons occupy the configuration of the normal ground state
 and a \={K} meson occupies the 0$s$ state. Of course, as done in the former section, 
we confirm after $J$ \& $T$ projections that 
the expectation values of $\hat{\mbox{\boldmath $J$}}$ and $\hat{\mbox{\boldmath $T$}}$ 
are equal to those we set beforehand. 
Fig. \ref{Energies} displays the 
behavior of the total binding energy and the decay width, 
from ppnK$^-$ to $^9$BK$^-$. The previous result \cite{AMDK} 
of $^8$BeK$^-$ is also shown there. 
According to the Table IV, all \={K} nuclei are bound by about 
100 MeV. Fig. \ref{Energies} shows us that 
the \={K} nuclei except for ppp{\Km} are bound below the ${\rm \Sigma}\pi$ threshold 
which is the main decay channel. 

In Fig. \ref{Energies}, the thresholds of various decay modes are also shown, 
from which we can estimate the stability of pppnK$^-$ to $^9$BK$^-$ 
for the strong interaction. 
First, pppnK$^-$ is found to be unstable for the strong interaction,  since its total binding energy (112.5 MeV) is smaller than that of ppnK$^-$ + p (117.8 MeV). 
On the other hand, 
the deepest thresholds for $^6$BeK$^-$, $^8$BeK$^-$ and $^9$BK$^-$ 
are ppnK$^-$+$^3$He (125.3 MeV), ppnK$^-$+$\alpha$+n (143.5 MeV) and 
$^8$BeK$^-$+p (159.0 MeV), while their total binding energies are 131.7 MeV, 
159.0 MeV and 173.5 MeV, respectively. 
Therefore, $^6$BeK$^-$, $^8$BeK$^-$ and $^9$BK$^-$ are 
stable for the strong interaction. 

As mentioned above, the K$^-$ meson is bound by about 100 MeV in all of the \={K} 
nuclei that we calculated. pK$^-$ is bound by 27 MeV based on the initial assertion 
that it forms $\Lambda(1405)$, and ppK$^-$ is bound by 48 MeV 
according to \cite{YA}. Therefore, the binding energy of the K$^-$ 
meson seems to be saturated in \={K} nuclei heavier than ppnK$^-$. 
We think that this saturation of the binding energy is related to the range of 
the \={K}N interaction. Since it is very short, the number of nucleons which 
a single K$^-$ meson can interact with is limited. We count up the nucleons 
staying in the region where the \={K} meson's density falls down 
from the maximum value $\rho^K_{MAX}$ to $\frac{1}{5} \; \rho^K_{MAX}$. 
In Table V, we show the number of strongly interacting nucleons (``Nucleon'') 
staying around the \={K} in various \={K} nuclei. 
Except for pppK$^-$, which has a peculiar structure, as mentioned  
in the later section, about 1.7 to 2.5 nucleons are found to stay near the \={K} meson. 

\begin{figure}
\begin{minipage}[t]{3.5cm}
   \includegraphics[width=1.00\textwidth]{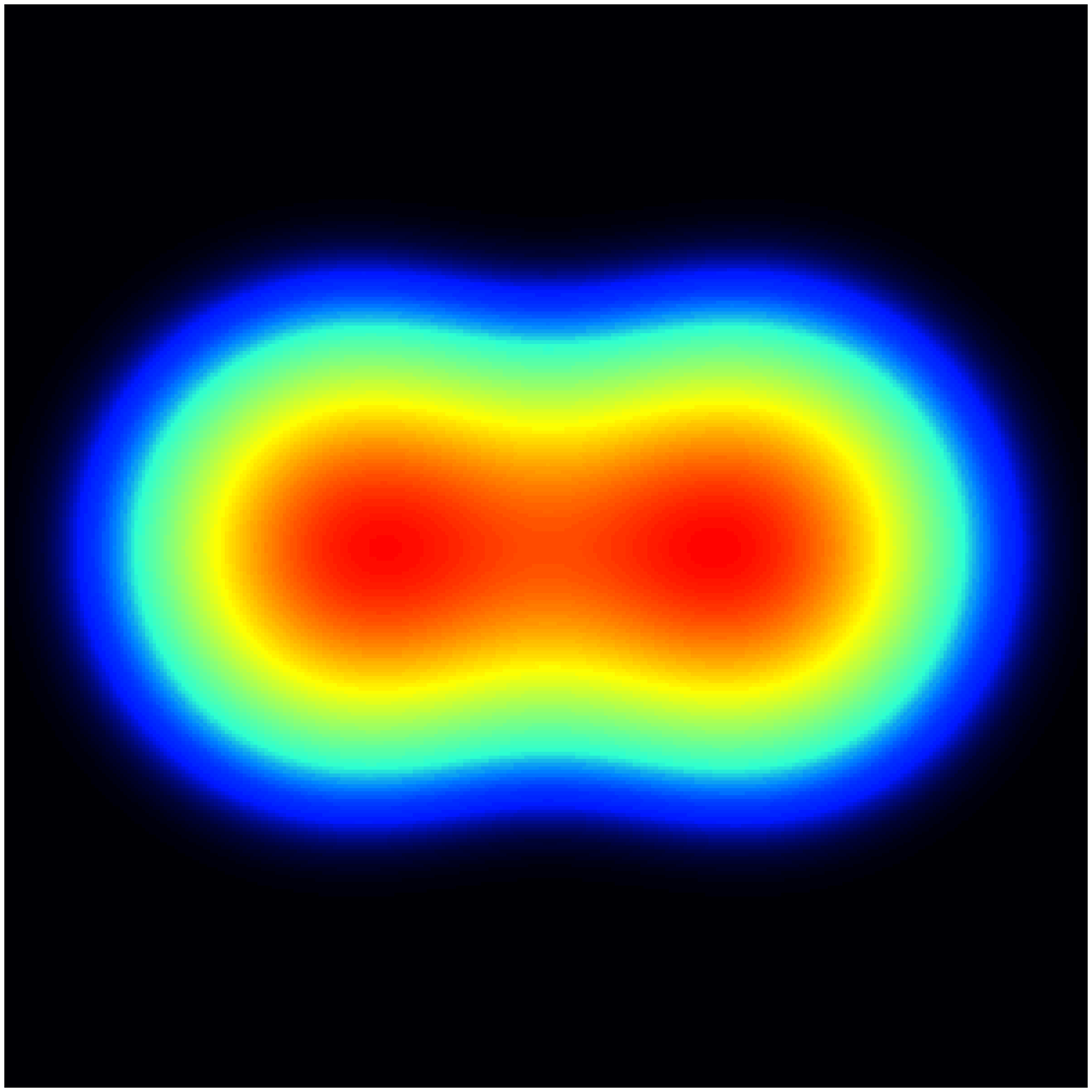}%

\end{minipage}
\hspace{0.1cm}
\begin{minipage}[t]{3.5cm}
   \includegraphics[width=1.00\textwidth]{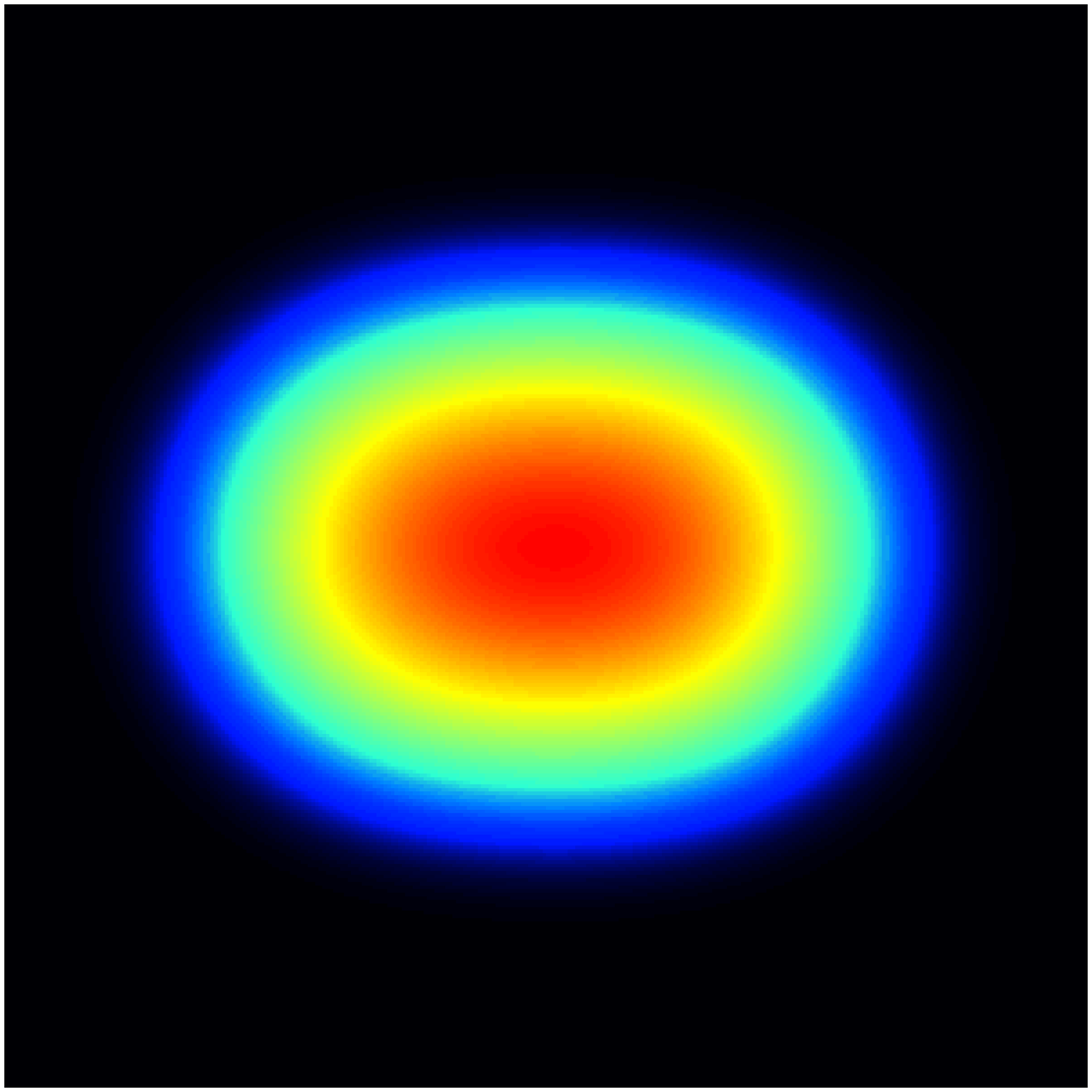}%

\end{minipage}

   \caption{\label{Dens6BeK}
Proton (left) and neutron (right) distributions of $^6$BeK$^-$.}
\end{figure}

\subsection{Density distribution}

Fig. \ref{Densities} displays the nucleon density distributions of 
ppnK$^-$, pppK$^-$, pppnK$^-$, $^6$BeK$^-$ and $^9$BK$^-$. 
It is found that \={K} nuclei have extremely dense and peculiar nucleon distributions. 

$^6$BeK$^-$ has a two-center-like structure similar to $^8$BeK$^-$ \cite{AMDK}. 
Fig. \ref{Dens6BeK} shows proton and neutron distributions separately. 
We can find that protons have a two-center-like structure and 
that neutrons stay between two pairs of protons, against our expectation 
that $^6$Be should have such a structure as $\alpha$+2p. 
We note that the \={K} meson's density distribution, which is not 
displayed here, is very similar to the neutrons' one. 
The structure of $^9$BK$^-$ is quite similar to that of $^8$BeK$^-$ \cite{AMDK}. 

\begin{figure}
\begin{minipage}[t]{3.6cm}
   \includegraphics[width=1.00\textwidth]{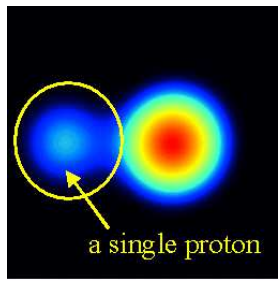}%
\end{minipage}

   \caption{\label{DenspppK}
Proton distribution in pppK$^-$.}
\end{figure}

The most exotic system of pppK$^-$ shows a very peculiar density distribution. 
Strictly speaking, this system has not only the component of pppK$^-$,  
but also that of ppn\={K}$^0$ due to the $I=0$ \={K}N interaction. 
We express this system as $^3_{\bar{\rm K}}$He. 
Fig. \ref{DenspppK} shows only proton density distribution extracted. 
We can see a ``satellite'' in this figure. 
Summing up the density of proton in the region of this satellite, 
we find that the proton number is nearly equal to one. Thus, this satellite is 
a single proton. In addition, the particle numbers of proton, 
neutron, K$^-$ and \={K}$^0$ are 2.67, 0.33, 0.67 and 0.33, respectively. 
We can understand these particle numbers and the density distribution 
consistently as follows.  
We regard this system as a single proton combining $^2_{\bar{\rm K}}$H, 
following its density distribution (Fig. \ref{DenspppK}). 
In isospin space,  
$^2_{\bar{\rm K}}$H means the $| T=\frac{1}{2}, T_z=\frac{1}{2} \rangle$ 
state, which is composed of two nucleons and a \={K} meson. 
The weight of each component included in it is determined by 
Clebsch-Gordan coefficients as shown below:  
\begin{widetext}
\begin{eqnarray}
| ^3_{\bar{\rm K}}{\rm He} \rangle & = & 
| {p} \rangle \otimes | ^2_{\bar{\rm K}}{\rm H} \rangle \;
= \; | {p} \rangle \otimes \left| T=\frac{1}{2}, T_z=\frac{1}{2} \right\rangle \nonumber \\
& = & 
| {p} \rangle \otimes 
\left( \sqrt{\frac{2}{3}} \; \left| T^N=1, T^N_z=1 ; T^K=\frac{1}{2}, T^K_z=-\frac{1}{2} \right\rangle 
\right.
- \left. \sqrt{\frac{1}{3}} \; \left| T^N=1, T^N_z=0 ; T^K=\frac{1}{2}, T^K_z=\frac{1}{2} \right\rangle \right) \nonumber \\
& = & | {p} \rangle \otimes \left( \sqrt{\frac{2}{3}} \; | {pp} \otimes {K}^- \rangle -  
\sqrt{\frac{1}{3}}  \; | {pn} \otimes \bar{K}^0 \rangle \right) \label{eq:pppK}
\end{eqnarray}
\end{widetext}

\noindent
The particle numbers counted according to Eq. (\ref{eq:pppK}) 
are quite the same as the above values.
After all, we come to an idea that in $^3_{\bar{\rm K}}$He 
one proton keeps its identity and that the residual part is 
composed of ppK$^-$ and pn\={K}$^0$, which are mixed due 
to the $I=0$ \={K}N interaction. Note, however, that 
the single proton is strongly bound to $^2_{\bar{\rm K}}$H. 
In addition, we studied the dependence of this system on the number of 
wave packets. Even if one nucleon is represented by  
four Gaussian wave packets (i.e. $N_n=4$ in Eq. (\ref{eq:N})), 
a proton is still pushed out, but a little less clearly. 
We consider that the third proton is pushed up to 
the 0$p$-shell due to Pauli Blocking, 
so that it forms a satellite-like structure. 
pppnK$^-$ has an extra neutron compared to pppK$^-$.
When a neutron is added to pppK$^-$, such a satellite-like structure disappears and 
a different structure is formed for pppnK$^-$.

\section{Summary \label{Summary}}

We improved the antisymmetrized molecular dynamics (AMD) 
regarding two points and applied it to systematic studies of \={K} nuclei. 
One of our improvements is ``pK$^-$/n\={K}$^0$ mixing'', which enables 
us to treat directly the coupling of pK$^-$ and n\={K}$^0$ through 
the $I=0$ \={K}N interaction. The other one is ``$J$ \& $T$ projections'' 
with which the strong isospin dependence of the \={K}N interaction 
is expected to be correctly treated. Thus, AMD is capable of treating 
the $I=0$ \={K}N interaction adequately,  
which plays an essential role in \={K} nuclei. 

We have investigated the properties of our new framework on ppnK$^-$.
After $J$ \& $T$ projections, the total system is found to possess 
the quantum numbers $J$ and $T$, 
which turn out to be equal to those which we had assigned beforehand. 
Namely, we have confirmed that the $J$ \& $T$ projections work correctly. 
The new result and the previous one are very similar to each other, but  
we can see the influence of pK$^-$/n\={K}$^0$ mixing 
in the density distributions of the protons and neutrons. 
Owing to the introduction of \={K}$^0$, 
the present wave function can form the eigen-state of isospin, 
i.e. $|^3_{\bar{\rm K}}{\rm H} (T=0) \rangle$. 
When we draw the proton and neutron distributions of 
$|^3_{\bar{\rm K}}{\rm H} (T=0) \rangle$, they are clearly 
different from those shown in our previous study. 
We confirm that the $|^3_{\bar{\rm K}}{\rm H} (T=0) \rangle$ is 
formed by the rotation of $|{\rm ppnK}^-\rangle$ in isospin space. 
Namely, $|{\rm ppnK}^-\rangle$ is considered to be an 
``intrinsic state in isospin space'' 
for the $|^3_{\bar{\rm K}}{\rm H} (T=0) \rangle$. 
This intrinsic state $|{\rm ppnK}^-\rangle$ is found to 
have quite the same proton and neutron distributions as those in our 
previous study. 
The $|^3_{\bar{\rm K}}{\rm H} (T=0) \rangle$ contains both $|{\rm pnn\bar{K}}^0\rangle$ 
and $|{\rm ppnK}^-\rangle$ components with the same ratio. 
This is consistent with the fact that 
the calculated proton and neutron numbers are both equal to 1.5.
The coupling of these two components due to the 
$I=0$ \={K}N interaction helps the binding of the total system 
$|^3_{\bar{\rm K}}{\rm H} (T=0) \rangle$.

We have studied ppnK$^-$, pppK$^-$, pppnK$^-$, $^6$BeK$^-$ and 
$^9$BK$^-$ with our new framework. 
All \={K} nuclei that we investigated are bound by about 100 MeV below the 
threshold of each nucleus+K$^-$. Except for pppK$^-$, they are bound 
below the ${\rm \Sigma}\pi$ threshold, which is the main decay channel. 
Since their decay width is 20 to 40 MeV and small 
compared to their binding energy, they appear to be discrete states.
For the strong interaction, pppnK$^-$ is found to be unstable, while 
$^6$BeK$^-$, $^8$BeK$^-$ and $^9$BK$^-$ are stable. 
We found that 
they have very different structures. Especially, pppK$^-$ shows 
an interesting satellite-like which is composed of a single proton. 
This proton keeps its identity and  is strongly bound by the main body. 

According to our present study, we predict various deeply bound \={K} nuclei. 
They have very peculiar structures with extremely high densities, 
which we have never seen. In the future, such \={K} nuclei as those investigated 
in the present paper may be explored experimentally. For instance, 
ppnK$^-$ can be formed from a $^4$He(stopped K$^-$, n) experiment, which 
is now under way at KEK \cite{Akaishi-Yamazaki,Iwasaki:01}. 
The use of in-flight (K$^-,N$) reactions is also proposed \cite{Kishimoto}. 
Proton-rich exotic \={K} nuclei are expected to be 
produced by (K$^-$, $\pi^-$) reactions via $\Lambda^*$ doorway states \cite{YA}.


Finally, we would like to mention possibility of the new framework of AMD. 
Since it can be extended straightforwardly to the case of multi \={K}'s, 
we can investigate multi-\={K} nuclei, which are closely related to 
kaon condensation and strange quark matter. 
The success of the new framework of AMD means that 
we can deal with even more fields 
of physics, because coupled-channel-like calculation can be carried out 
in the new version of AMD. For example, we can study 
$\Lambda$-$\Sigma$ mixing in hypernuclei \cite{LamSigmix} 
with this new framework. 


\section{Acknowledgment}

One of the authors (A.D.) thanks 
Dr. M. Kimura for giving him the graphic tools he made, 
and Dr. Y. Kanada-En'yo for fruitful discussions on our new framework. 
A part of calculations in this paper is made using 
Scalar System Alpha1 in Yukawa Institute for Theoretical Physics. 
This work is supported by Grant-in-Aid for Scientific Research of 
Monbukagakusho of Japan.

\end{document}